\documentclass{aa}

\usepackage{epsfig}

\newcommand{\lesssim}{\ensuremath{\stackrel{<}{\sim}}}
\newcommand{\xten}[1]{\ensuremath{\times 10^{#1}}}
\newcommand{\ten}[1]{\ensuremath{10^{#1}}}
\newcommand{\Hmol}{\ensuremath{\mathrm{H}_2}}
\newcommand{\water}{\ensuremath{\mathrm{H_2O}}}
\newcommand{\HA}{\ensuremath{\mathrm{H}\alpha}}
\newcommand{\PA}{\ensuremath{\mathrm{Pa}\alpha}}
\newcommand{\BA}{\ensuremath{\mathrm{Br}\alpha}}
\newcommand{\BG}{\ensuremath{\mathrm{Br}\gamma}}
\newcommand{\HNU}{\ensuremath{\mathrm{h}\nu}}
\newcommand{\AV}{\ensuremath{A_\mathrm{V}}}
\newcommand{\AWL}{\ensuremath{A_\lambda}}
\newcommand{\Lo}{\ensuremath{\mathrm{L}_\odot}}
\newcommand{\Mo}{\ensuremath{\mathrm{M}_\odot}}
\newcommand{\Zo}{\ensuremath{\mathrm{Z}_\odot}}

\newcommand{\LX}{\ensuremath{L_\mathrm{X}}}
\newcommand{\LFIR}{\ensuremath{L_\mathrm{FIR}}}

\newcommand{\QH}{\ensuremath{Q(\mathrm{H})}}
\newcommand{\NH}{\ensuremath{N_\mathrm{H}}}
\newcommand{\EV}{\ensuremath{\,\mathrm{ev}}}
\newcommand{\KEV}{\ensuremath{\,\mathrm{keV}}}
\newcommand{\ERG}{\ensuremath{\,\mathrm{erg}}}
\newcommand{\K}{\ensuremath{\,\mathrm{K}}}
\renewcommand{\S}{\ensuremath{\,\mathrm{s}}}
\newcommand{\YR}{\ensuremath{\,\mathrm{yr}}}
\newcommand{\MPC}{\ensuremath{\,\mathrm{Mpc}}}
\newcommand{\PC}{\ensuremath{\,\mathrm{pc}}}

\newcommand{\CM}{\ensuremath{\,\mathrm{cm}}}
\newcommand{\MIC}{\ensuremath{\,\mathrm{\mu m}}}
\newcommand{\MAG}{\ensuremath{\,\mathrm{mag}}}
\newcommand{\1}{\ensuremath{^{-1}}}
\newcommand{\2}{\ensuremath{^{-2}}}
\newcommand{\forb}[2]{\ensuremath{[\mathrm{#1}\,\textsc{\lowercase{#2}}]}}
\newcommand{\perm}[2]{\ensuremath{\mathrm{#1}\,\textsc{\lowercase{#2}}}}
\newcommand{\WL}{\ensuremath{\lambda}}
\newcommand{\OIII}{\forb{O}{III}}
\newcommand{\NeV}{\forb{Ne}{V}}
\newcommand{\NeII}{\forb{Ne}{II}}
\newcommand{\FeII}{\forb{Fe}{II}}
\newcommand{\NII}{\forb{N}{II}}
\newcommand{\HeI}{\perm{He}{I}}

\begin{document}

\thesaurus{11.01.2 -- 11.09.1 -- 11.14.1 -- 11.19.1 -- 11.19.3 -- 13.09.1}

\title{The Elusive Active Nucleus of NGC 4945.
\thanks{Based on observations made with the NASA/ESA Hubble Space Telescope,
obtained at the Space Telescope Science Institute, which
is operated by AURA, Inc., under NASA contract NAS 5--26555.
Also based on observation collected at European Southern Observatory,
La Silla, Chile.} }

\author{A. Marconi\inst{1},
	E. Oliva\inst{1},
	P.P. van der Werf\inst{2},
	R. Maiolino\inst{1},
	E.J. Schreier\inst{3},
	F. Macchetto\inst{3,4},
	\and A.F.M. Moorwood\inst{5}
	}

\offprints{A. Marconi}

\institute{
	  Osservatorio Astrofisico di Arcetri,
	  Largo E. Fermi 5, I-50125 Firenze, ITALY
	  \and
	  Sterrewacht Leiden, P.O. Box 9513, 2300 RA Leiden, The Netherlands
	  \and
	  Space Telescope Science Institute
	  3700 San Martin Drive, Baltimore, MD 21218, USA
	  \and
	  Affiliated to ESA science division
	  \and
	  European Southern Observatory
	  Karl-Schwarzschild-Strasse 2, 85748 Garching bei M\"unchen, Germany
}

\date{Received ... Accepted ... }

\authorrunning{Marconi et al.}
\titlerunning{The elusive AGN of NGC 4945}
\maketitle

\begin{abstract}

We present new HST NICMOS observations of NGC 4945, a starburst 
galaxy hosting a highly obscured active nucleus that is
one of the brightest extragalactic sources at 100 keV.
The HST data are complemented with ground based \FeII\
line and mid--IR observations. 

A 100pc-scale starburst ring is detected in \PA, while \Hmol\ traces
the walls of a super bubble opened by supernova-driven winds.
The conically shaped cavity is particularly prominent in \PA\
equivalent width and in the \PA/\Hmol\ ratio.
Continuum images are heavily affected by dust extinction and the
nucleus of the galaxy is located
in a highly reddened region with an elongated, disk-like morphology.
No manifestation of the active nucleus is found,
neither a strong point source nor dilution in CO stellar features,
which are expected tracers of AGN activity.

Even if no AGN traces are detected in the near-IR,
with the currently available data it is still not possible
to establish whether the bolometric luminosity of the object is
powered by the AGN or by the starburst: we demonstrate
that the two scenarios constitute equally viable alternatives.
However, the absence of any signature other than in the hard X-rays 
implies that, in both scenarios, the AGN is non-standard: 
if it dominates, it must be obscured in all directions,
conversely, if the starburst dominates, the AGN must lack UV photons
with respect to X-rays.

An important conclusion is that
powerful AGNs can be hidden even at mid-infrared
wavelengths and, therefore, the nature of luminous dusty galaxies
cannot be always characterized by long-wavelength data alone
but must be complemented with sensitive hard X-ray observations.


\end{abstract}

\section{Introduction}

A key problem in studies of objects emitting most of their energy
in the FIR/submm is to establish the
relative importance of highly obscured
Active Galactic Nuclei (AGN) and starburst activity.
In particular, it is important to know if it is still
possible to hide an AGN, contributing significantly to the bolometric
emission, when optical to mid-IR spectroscopy and imaging  
reveal only a starburst component.

Several pieces of evidence suggest that most cosmic
AGN activity is obscured.
Most, and possibly all, cores of large galaxies host a
supermassive black hole (\ten{6}--\ten{9}\Mo;
e.g. Richstone et al. \cite{richstone}).
To complete the formation process in a Hubble time,
accretion must proceed at high rates,
producing quasar luminosities ($L\sim\ten{12}\Lo$).
However the observed black hole density is an order of magnitude
greater than that expected from the observed quasar light,
assuming accretion efficiency of 10\%,
suggesting that most of the accretion history is obscured
(e.g. Fabian \& Iwasawa \cite{fabian99}, and references therein).
It is estimated either that 85\% of all AGNs are obscured (type 2)
or that 85\% of the accretion history of an object is hidden from view.

In addition, the hard X-ray background ($>1\KEV$)
requires a large population of obscured AGNs
at higher redshifts ($z\sim1$) since the
observed spectral energy distribution cannot
be explained with the continua of Quasars, 
i.e. un--obscured (type 1) AGNs (Comastri et al. \cite{comastri},
Gilli et al. \cite{gilli99}).
Despite the above evidence, detections of obscured AGNs at
cosmological distances are still sparse (e.g. Akiyama et al. \cite{akiyama}).

Ultra Luminous Infrared Galaxies (ULIRGs; see
Sanders \& Mirabel \cite{sanders96} for a review)
and the sources detected in recent far-infrared and submm surveys 
performed with ISO and SCUBA (e.g. Rowan-Robinson et al. \cite{rowanrob},
Blain et al. \cite{blain} and references therein)
are candidate to host the missing population of type 2 AGNs.
However, mid-IR ISO spectroscopy
has recently shown that ULIRGs are mostly powered
by starbursts and that no trace of AGNs is found in the majority of cases
(Genzel et al. \cite{genzel98}; Lutz et al. \cite{lutz98}).
Yet, the emission of a hidden AGN could be heavily absorbed
even in the mid-IR. Indeed, the obscuration of the AGN could be related
to the starburst phenomenon, as observed for Seyfert 2s
(Maiolino et al. \cite{maiolino95}).
Fabian et al. (\cite{fabian98}) proposed
that the energy input from supernovae and stellar winds prevents 
interstellar clouds from collapsing into a thin disk, thus maintaining them 
in orbits that intercept the majority of the lines of sight from 
an active nucleus.

In this paper, we investigate the existence
of completely obscured AGNs and the Starburst-AGN connection through
observations of NGC 4945,  one of the closest galaxies where
an AGN and starburst coexist.
NGC 4945 is an edge-on ($i\sim 80^\circ$), nearby ($D=3.7\MPC$) 
SB spiral galaxy hosting a
powerful nuclear starburst (Koornneef \cite{koorn}; 
Moorwood \& Oliva \cite{moorwood94a}).
It is a member
of the Centaurus group and, like the more famous Centaurus A (NGC 5128),
its optical image is marked by dust extinction
in the nuclear regions. The ONLY evidence for a hidden AGN comes from the
hard X-rays where NGC 4945 is characterized by a Compton-thick spectrum
(with an absorbing column density of $\NH=5\xten{24}\CM\2$,
Iwasawa et al. \cite{iwasawa93})
and one of the brightest 100\KEV\ emissions among extragalactic sources
(Done et al. \cite{done96}).
Recently, BeppoSAX clearly detected variability in the 13-200\KEV\ band
(Guainazzi et al., \cite{guainazzi}).

Its total infrared luminosity derived from IRAS data
is $\sim 2.4\xten{10}\Lo$ (Rice et al. \cite{rice88}),
$\sim 75\%$ of which arises from
a region of $\le 12\arcsec\times9\arcsec$ centered on
the nucleus (Brock et al. \cite{brock88}).
Although its star formation and supernova rates are moderate,
$\sim 0.4\,\Mo\YR\1$
and $\sim 0.05\YR\1$ (Moorwood \& Oliva \cite{moorwood94a}),
the starburst activity is concentrated in the central $\sim 100\PC$ and
has spectacular consequences
on the circumnuclear region which is characterized by a conical cavity
evacuated by a supernova-driven wind (Moorwood et al. \cite{moorwood96a}).  

The radio emission is characterized by a compact non-thermal core
with a luminosity of $\simeq 8\xten{38}\ERG\S\1$
(Elmouttie et al. \cite{elmouttie}).
It is one of the first H$_2$O and OH megamaser sources detected
(dos Santos \& Lepine \cite{dossantos}; Baan \cite{baan85}) and
the H$_2$O maser was mapped by Greenhill et al. (\cite{greenhill})
who found the emission linearly
distributed along the position angle of the galactic disk and with a 
velocity pattern suggesting the presence of a 
$\sim\ten{6}\Mo$ black hole.
Mauersberger et al. (\cite{mauersberger}) mapped the $J=3-2$ line of $^{12}$CO
which is mostly concentrated within the nuclear $\sim 200\PC$.

We present new line and continuum images obtained with the
{\it Near Infrared Camera and Multi
Object Spectrograph} (NICMOS) on-board the Hubble Space Telescope (HST),
aimed at detecting AGN activity in the near-infrared.
These observations are complemented by recent ground based near- and
mid-IR observations obtained at the European Southern Observatory.
Section \ref{sec:obs} describes the observations and data reduction techniques.
Results are presented in Section \ref{sec:res} and discussed
in Section \ref{sec:discuss}.
Finally, conclusions will be drawn in Sec. \ref{sec:conclus}.
Throughout the paper we assume a distance of 3.7\MPC\
(Mauersberger et al. \cite{mauersberger}),
whence 1\arcsec\ corresponds to $\simeq18$\PC.

\begin{figure*}
\begin{center}
\begin{tabular}{cc}
 \epsfig{figure=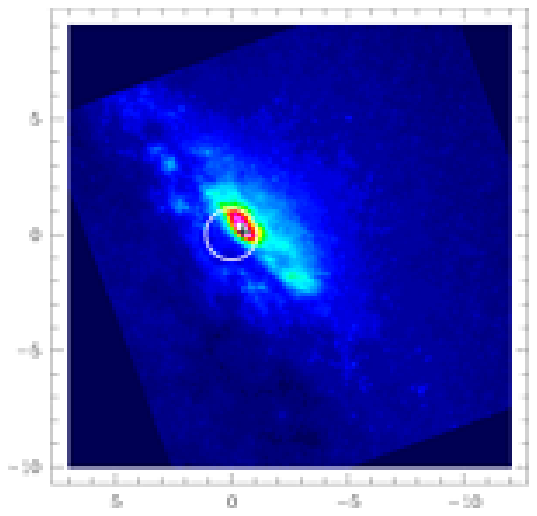,width=0.45\linewidth} &
 \epsfig{figure=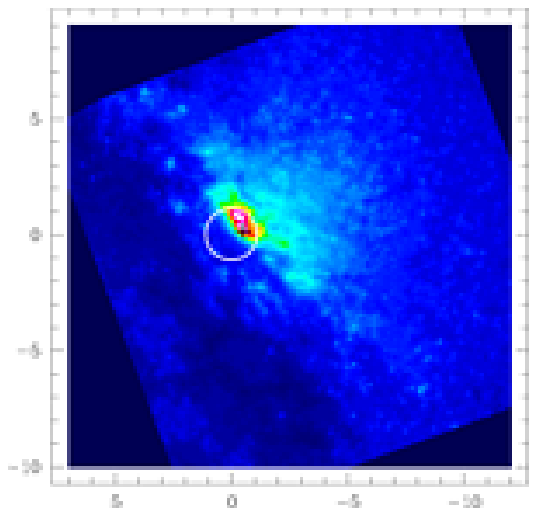,width=0.45\linewidth} \\
 & \\
 \epsfig{figure=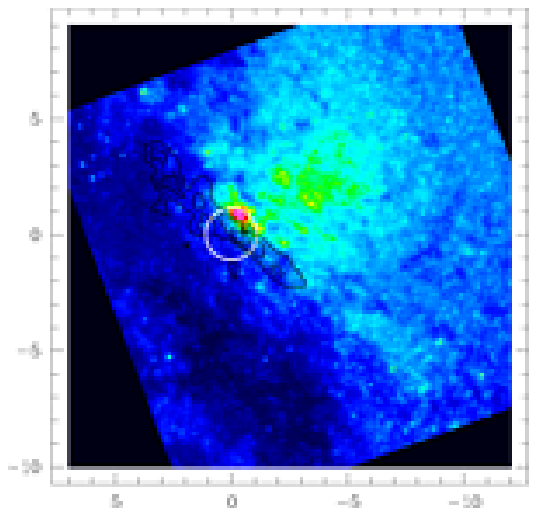,width=0.45\linewidth} &
 \epsfig{figure=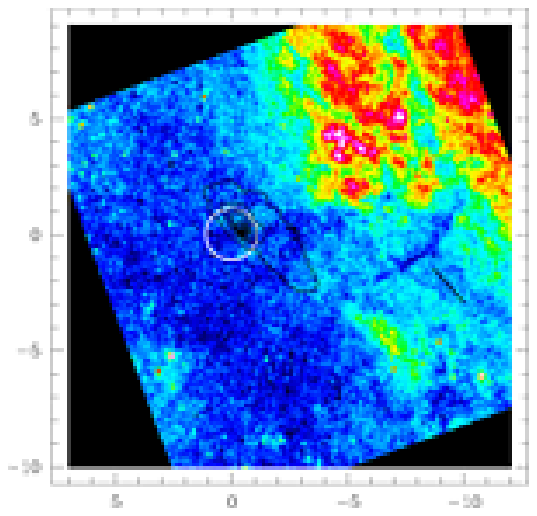,width=0.45\linewidth} \\
 & \\
 \epsfig{figure=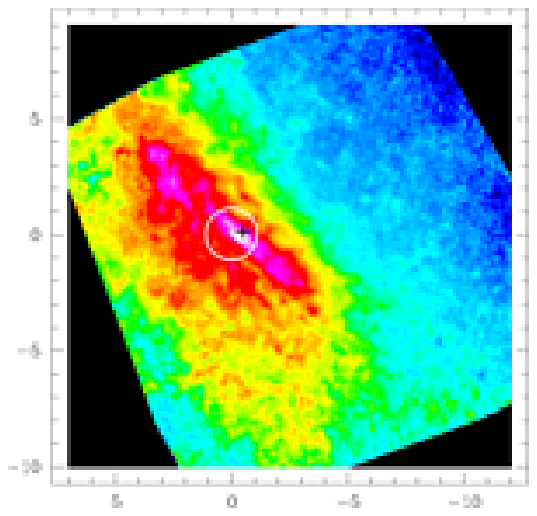,width=0.45\linewidth}    &
 \hspace{14pt}\vspace*{-14pt}
 \epsfig{figure=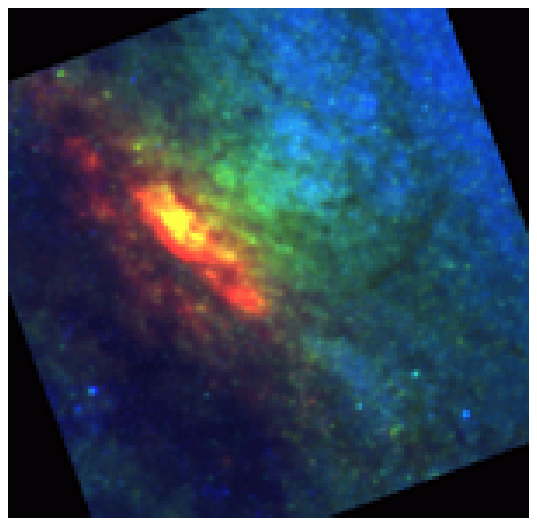,width=0.38\linewidth} \\
\end{tabular}
\end{center}
\caption{\label{fig:cont}
(a) F222M image (K band).
North is up and East is left.
The cross marks the location of the K nucleus and
the circle represents the uncertainty on the position of the H$_2$O maser
given by Greenhill et al. (1997).
Units of the frame box are seconds of arc.
The origin is at the nominal location of the H$_2$O maser.
(b) F160W image (H). Notation as in panel (a).
(c) F110W image (J). Notation as in panel (a).
The black contours are from the H-K color image at 1.8, 2 and 2.2 levels.
(d) F606W image (R band). Notation as in panel (a) except for the contours
which are from the K band image. (e) H-K image. Symbols are as in (a).
(f) Truecolor (Red=F222M, Green=F110W, Blue=F606W) image.
}
\end{figure*}

\begin{figure*}[!]
\begin{center}
\begin{tabular}{cc}
 \epsfig{figure=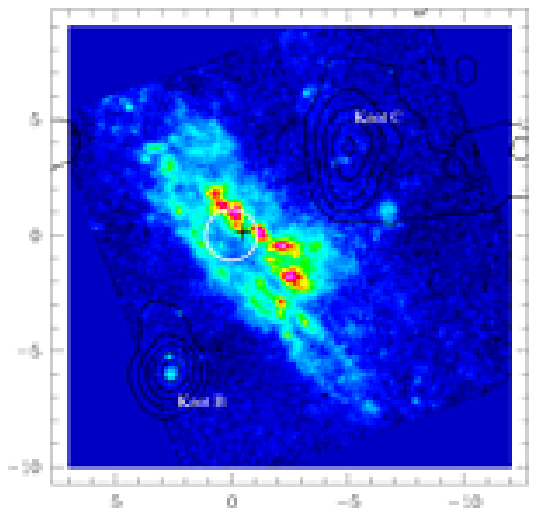,width=0.45\linewidth} &
 \epsfig{figure=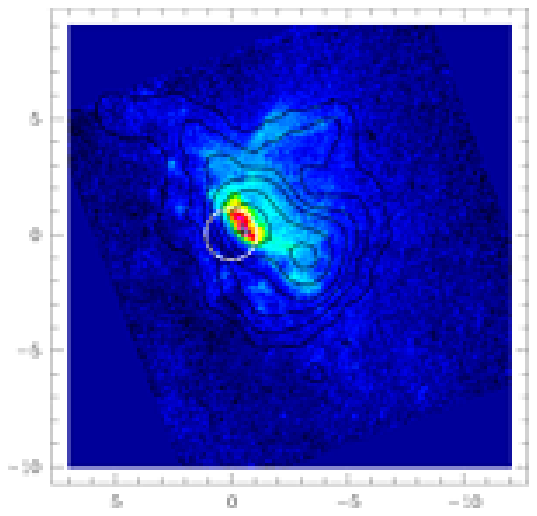,width=0.45\linewidth} \\
 & \\
 \epsfig{figure=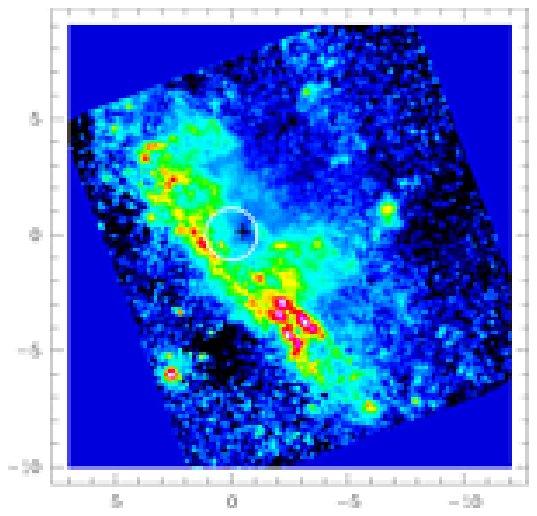,width=0.45\linewidth} &
 \epsfig{figure=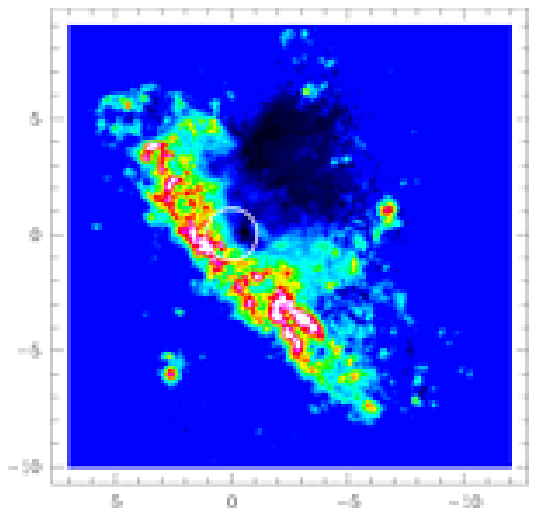,width=0.45\linewidth} \\
 & \\
 \epsfig{figure=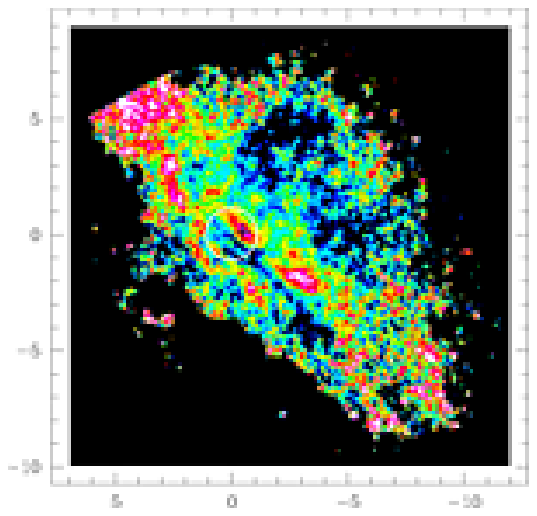,width=0.45\linewidth} &
 \hspace{14pt}\vspace*{-14pt}
 \epsfig{figure=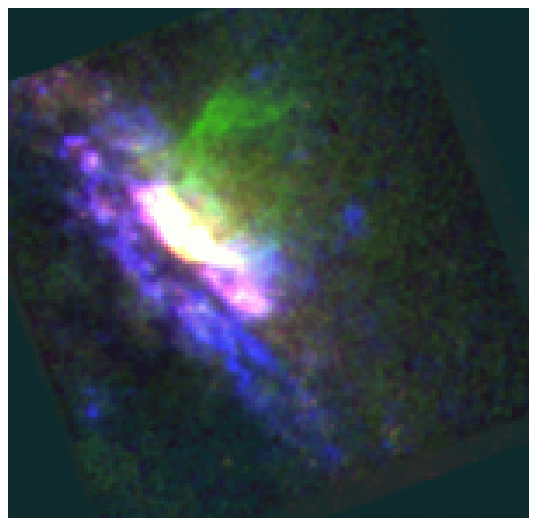,width=0.38\linewidth} \\
\end{tabular}
\end{center}
\caption{\label{fig:line}
(a) Pa$\alpha$ image. Symbols as in Fig. 1.
The black contours are from the \HA+\NII\ image by Moorwood et al. (1996).
(b) H$_2$ image. Black contours are from the blue ground-based \FeII\ image,
(c) Equivalent width of Pa$\alpha$.
(d) \PA/\Hmol\ image. Symbols as in Fig. 1.
(e) CO index. Symbols as in Fig. 1.
(f) Truecolor line image (Red=F222M, Green=\Hmol, Blue=\PA) image.
}
\end{figure*}

\section{\label{sec:obs} Observations and Data Reduction}

\begin{table}
\caption{\label{tab:log} Log of HST observations.}
\begin{tabular}{lccl}
\hline\hline
Dataset & Filter & T$_{exp}$ (sec) & Description \\
\hline
\\
n4mq01010 & F110W & 768	   & J 			  \\
n4mq02010 & F160W & 768	   & H 			  \\
n4mq01040 & F222M & 288    & K 			  \\
n4mq01070 & F237M & 288    & CO 		  \\
n4mqa1010 & F222M & 288    & K background 	  \\
n4mqa1020 & F237M & 288    & CO background 	  \\
\\
n4mqb1nrq & F187N & 320    & \PA\		  \\
n4mqb1nuq & F190N & 320    & \PA\ continuum	  \\
n4mqb1o0q & F190N & 320    & \PA\ continuum	  \\
n4mqb1o3q & F187N & 320    & \PA\		  \\
n4mqb1o9q & F187N & 320    & \PA\		  \\
n4mqb1odq & F190N & 320    & \PA\ continuum	  \\
n4mqb1oiq & F190N & 320    & \PA\ continuum	  \\
n4mqb1olq & F187N & 320    & \PA\		  \\
\\
n4mqb1opq & F212N & 320    & \Hmol\		  \\
n4mqb1osq & F190N & 320    & \Hmol\ continuum	  \\
n4mqb1ovq & F190N & 320    & \Hmol\ continuum	  \\
n4mqb1ozq & F212N & 320    & \Hmol\		  \\
n4mqb1p2q & F212N & 320    & \Hmol\		  \\
n4mqb1p5q & F190N & 320    & \Hmol\ continuum	  \\
n4mqb1p9q & F190N & 320    & \Hmol\ continuum	  \\
n4mqb1pcq & F212N & 320    & \Hmol\		  \\
\\
u29r2p01t & F606W & 80	   & R archive	  \\
u29r2p02t & F606W & 80     & R archive       \\
u2e67z01t & F606W & 500    & R archive       \\
\\
\hline
\end{tabular}
\end{table}

\begin{table*}
\caption{\label{tab:loggr}Log of Ground Based observations.}
\begin{tabular}{lcccl}
\hline\hline
Image      & T$_{exp}$ (min) & Date          & Instrument & Telescope\\
\hline
\\
L$^\prime$ & 10              & May 30, 1996  & IRAC1      & ESO/MPI 2.2m \\
N          & 40              & May 27, 1996  & TIMMI      & ESO 3.6m \\
\FeII\     & 24              & April 1, 1998 & IRAC2B     & ESO/MPI 2.2m \\
\\
\hline
\end{tabular}
\end{table*}

The nuclear region of NGC 4945 was observed on March 17$^{th}$ and
25$^{th}$, 1998, with NICMOS Camera 2 (MacKenty et al. \cite{mackenty})
using narrow and broad band filters for imaging in lines and continuum.
HST observations are logged in Tab. \ref{tab:log}.
All observations were carried out with a MULTIACCUM sequence (MacKenty et al.
\cite{mackenty})
and the detector was read out non-destructively several times
during each integration to remove cosmic rays hits and correct
saturated pixels.
For each filter we obtained several exposures with the object 
shifted by $\sim 1\arcsec$ on the detector to remove bad pixels.
The observations in the F222M and F237M filters
were also repeated on a blank sky area several arcminutes away
from the source to remove thermal background emission.
For narrow band images, we obtained subsequent
exposures in line and near continuum filters with the object
at several positions on the detector.

The data were re-calibrated using the pipeline software CALNICA v3.2
(Bushouse et al. \cite{bushouse}). 
A small (few percent) drift in the NICMOS bias level caused an error
in the flat-fielding procedure which resulted in spurious artifacts
in the final images (the so-called "pedestal problem" -- Skinner,
Bergeron \& Daou \cite{skinner}). Given the strong signal from the galaxy,
such artifacts are only visible in ratio or difference images.
This effect was effectively removed using the pedestal
estimation and quadrant equalization software developed by Roeland
P. van der Marel which subtracts a constant bias
level times the flat-field, minimizing the standard deviation
in the images. For each filter, the corrected
images were then aligned via cross-correlation and combined.
Flux calibration of the images was achieved by multiplying the count rates
(adu\S\1) for the PHOTFLAM (\ERG\CM\2\,adu\1) conversion factors
(MacKenty et al. \cite{mackenty}).

The narrow band images obtained at wavelengths adjacent to the
\PA\ and \Hmol\ lines where used for continuum subtraction.
The procedure was verified by rescaling
the continuum by up to $\pm 10\%$ before subtraction and establishing that this did not significantly affect the observed emission-line structure.

WFPC2 observations in the F606W (R band) filter were retrieved from the
Hubble Data Archive and\\ re-calibrated with the
standard pipeline software (Biretta et al. \cite{biretta}).

Ground-based observations were obtained at the European Southern
Observatory at La Silla (Chile) in the continuum L$^\prime$ (3.8\MIC) and
N (10\MIC) bands, and in the \FeII\ 1.64\MIC\ emission line and
are logged in Tab. \ref{tab:loggr}.  The L$^\prime$ 
image was obtained with IRAC1 (Moorwood et al. \cite{moorwood94b})
at the ESO/MPI 2.2\,m telescope on May 30, 1996 using an SBRC 58$\times$62 pixel
InSb array with a pixel size of 0\farcs45. Double beam-switching was used,
chopping the telescope secondary mirror every 0.24\S\ and nodding the
telescope every 24\S\ to build up a total on-source integration time of
10\,minutes in a seeing of 0\farcs9. The N-band image was obtained with
TIMMI (K\"aufl et al. \cite{kaufl}) at the ESO 3.6m telescope on May 27, 1996
using a 64$\times$64 Si:Ga array with 0\farcs46 pixels. Again using
double beam switching, total on-source integration time was 40 minutes
in 1\arcsec\ seeing. The \FeII\ 1.64\MIC\ image was taken with with the
IRAC2B camera (Moorwood et al. \cite{moorwood92})
on the ESO/MPI 2.2m telescope on April 1, 1998, using a
256$\times$256 Rockwell NICMOS3 HgCdTe array with 0\farcs51 pixels. The
\FeII\ line was scanned with a $\lambda/\Delta\lambda=1500$ Fabry-Perot
etalon covering three independent wavelength settings on the line and
two on the continuum on either side of the line, for a total integration
times of 24\,minutes on the line in 0\farcs9 seeing. Standard procedures
were used for sky subtraction, flat fielding, interpolation of hot and
cold pixels at fixed positions on the array, recentering and averaging
of the data. For the \FeII\ data, the continuum was determined from the
two off-line channels and subtracted from the on-line data.
The integrated \FeII\ line flux is in excellent agreement with the value
determined by Moorwood and Oliva (\cite{moorwood94a}).

\section{\label{sec:res}Results}

\begin{figure*}
\begin{center}
\begin{tabular}{cc}
 \epsfig{figure=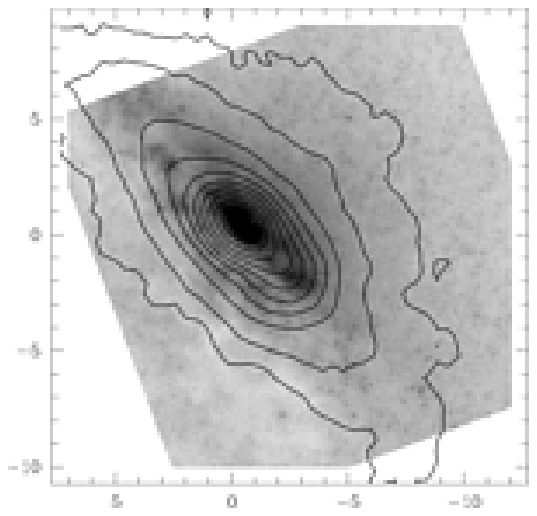,width=0.48\linewidth} &
 \epsfig{figure=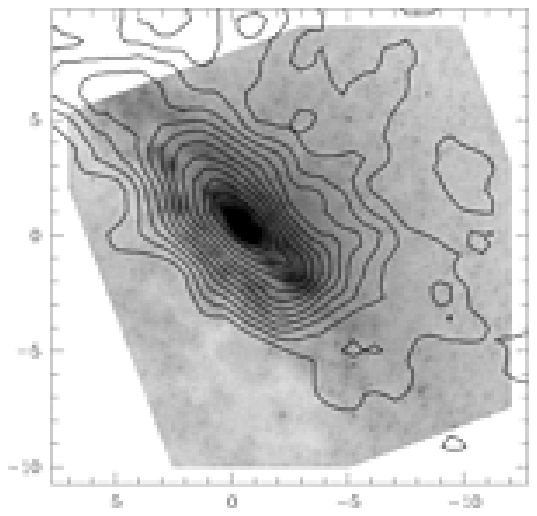,width=0.48\linewidth} \\
\end{tabular}
\end{center}
\caption{\label{fig:midir}
(a) L band contours overlayed on the NICMOS K band image
(displayed with a logarithmic look--up table).
Contours are 0.005 and from 0.01 to 0.14 with step of 0.01
(units of \ten{-16}\ERG\CM\2\S\1\AA\1). The frame boxes are
centered on the nucleus, identified with the \Hmol\ maser position.
(b) N band contours overlayed on the NICMOS K band image.
Contours are from 0.02 to 0.08 with step of 0.003
(units and notation as above). }
\end{figure*}

\subsection{Morphology}

Panels a--d in Figure \ref{fig:cont}
are the continuum images in the NICMOS
K, H, J and WFPC2 R filters\footnote{Color images are also available at
http://www.arcetri.astro.it/$\sim$marconi}.
The cross marks the position of the K band peak and the
circle is the position of the H$_2$O maser measured by
Greenhill et al. (\cite{greenhill}). The radius of the circle is the
$\pm1$\arcsec\ r.m.s. uncertainty of the astrometry performed on the
images and based on the Guide Star Catalogue (Voit et al. \cite{voit}).
The position of the K peak is offset by $\sim 0\farcs5$
from the location of the H$_2$O maser, hereafter identified with
the location of the nucleus of the galaxy. Note that 
this offset is still within the absolute astrometric uncertainties
of the GSC and the K peak could be coincident with the nucleus.
The continuum images are also shown with a "true color" RGB representation
in Fig. \ref{fig:cont}f (Red=F222M, Green=F110W, Blue=F606W).
A comparison of photometry between our data and earlier published
results is not straightforward since the NICMOS filters are different 
from the ones commonly used.
However, as shown in table \ref{tab:comp}, our measured fluxes in
6\arcsec\ and 18\arcsec\ circular apertures centered on the K band peak
are within 15-30\%\ of the ones by Moorwood \& Glass (\cite{moorwood84})
measured in the same areas.

\begin{table}
\caption{\label{tab:comp}Comparison with ground based photometric data.}
\begin{tabular}{ccccc}
\hline\hline
\\
	& 	& \multicolumn{2}{c}{Moorwood \& Glass 1984} & This work \\
Band 	& $\O$ (\arcsec) & mag	& Flux$^a$	& Flux$^a$		\\
\\
\hline
\\
K	& 6\arcsec  & 9.34	& 7.1\xten{-15} & 9.5\xten{-15}	\\
K	& 18\arcsec & 8.12	& 2.2\xten{-14} & 2.5\xten{-14}	\\
H	& 6\arcsec  & 10.7	& 5.7\xten{-15} & 7.1\xten{-15}	\\
H	& 18\arcsec & 9.15	& 2.4\xten{-14} & 2.7\xten{-14}	\\
J	& 6\arcsec  & 12.70    & 2.4\xten{-15}	& 2.7\xten{-15}	\\
J	& 18\arcsec & 10.80    & 1.4\xten{-14}	& 1.6\xten{-14}	\\
\\
\hline\hline
\end{tabular}
\\
$^a$ In units of \ERG\S\1\CM\2\AA.\\
$^b$ The K band of ground based observations corresponds to the F222M NICMOS
filter. Similarly H and J corresponds to F160W and F110W, respectively.
\end{table}

Figure \ref{fig:cont}e is the H-K color map. H-K contours are also 
overlayed on the J image in Fig. \ref{fig:cont}c.
At the location of the maser, South-East of the K band peak,
emission from galactic stars is obscured by a dust lane oriented 
along the major axis of the galactic disk.
The morphology of this resembles an edge-on disk with a 4\farcs5
radius (80\PC) and probably marks the region where high density 
molecular material detected in CO
is concentrated (e.g. Mauersberger et al. \cite{mauersberger}).
The average H-K color is $\sim 1.7$ with a peak value of 2.3.
The dust lane has a sharp southern edge which is not very evident in the color map and can be explained by
saturated absorption:
background H and K emission becomes completely undetectable and the color
is dominated by foreground stars. Therefore, the sharp K edge 
is evidence for a region with such high extinction that it is not detected
even in the near-IR.
With this dust distribution, 
the observed morphology in the continuum images is the result of
an extinction gradient in the direction perpendicular to the
galactic disk. Patchy extinction is also
present all over the field of view.
At shorter wavelengths, the morphology is more irregular
because dust extinction is more effective (the same effect seen so obviously in Centaurus A, cf. Schreier et al. \cite{schreier98},
Marconi et al. \cite{marconi99}) and the
conical cavity extensively mapped by Moorwood et al. (\cite{moorwood96a})
becomes more prominent:  there, the
dust has been swept away by supernova-driven winds.
Indeed, in the R image, significant emission is detected only in the wind-blown
cavity which presents a clear conical morphology 
with well defined edges and apex lying $\simeq 3\arcsec$
from the K peak. Due to the above mentioned
reddening gradient, the apex of the cone gets closer to the
nucleus with increasing wavelength
(compare with R band and Pa$\alpha$/H$_2$ images -- see below).

The continuum-subtracted Paschen $\alpha$ image
(Fig. \ref{fig:line}a) shows the presence of several strong emission line knots 
along the galactic plane, very likely resulting from a circumnuclear 
ring of star formation seen almost edge-on.
"Knot B" of Moorwood et al. (\cite{moorwood96a}) is clearly observed 
South East of the nucleus while "Knot C", North-West 
of the nucleus, is barely detected.
Both knots are also marked on the figure.

Dust extinction strongly affects the \PA\ morphology
making very difficult to trace the ring and locate its center; 
a likely consequence is 
the apparent misalignment between the galaxy nucleus and the ring center.
The observed ring of star formation is similar to what has been found in other
starburst galaxies (cf. Moorwood \cite{moorwood96b}).
The starburst ring could result from two alternative scenarios:
either the starburst originates at the nucleus,
and then propagates outward forming a ring
in the galactic disk; or the
ring corresponds to the position of
the inner Lindblad resonance  
where the gas density is naturally increased
by flow from both sides (see the review in Moorwood \cite{moorwood96b}).

Panel b shows the continuum-subtracted H$_2$ image which 
traces the edges of the wind-blown cavity. As expected, the morphology is 
completely different from that of \PA\ which traces mainly starburst activity.
Note the strong H$_2$ emission close to the nucleus at the
apex of the cavity with an elongated, arc-like morphology.
The \Hmol\ flux in a $6\arcsec\times 6\arcsec$ aperture centered on the K 
band peak is 1.1\xten{-13}\ERG\CM\2\S\1\ and corresponds to $\sim 70\%$ 
of the total integrated emission in the NICMOS field of view.
This is in good agreement with the 1.29$\pm 0.05$ found by
Koornneef \& Israel (\cite{koorn96}) in an equally sized aperture
and the integrated 3.1\xten{-13}\ERG\CM\2\S\1\ from the map by
Moorwood \& Oliva (\cite{moorwood94a}).
We remark that contamination of \Hmol\ emission by \HeI\WL 2.112\MIC\
is unlikely since the line was detected neither by Koornneef
(\cite{koorn}) nor by Moorwood \& Oliva (\cite{moorwood94a}) and
from their spectra we can set an upper limit
of 5-10\%\ to the \HeI/\Hmol\ ratio.

Panel c in Figure \ref{fig:line} shows that the  equivalent
width of \PA\ is up to 150--200\AA\ in the star forming regions,
but much lower in the wind-blown cavity.
Since near-IR continuum emission within the cone is not significantly
higher than in the surrounding medium,
the low equivalent width within the cone is due
to weaker \PA\ emission, the likely consequence of low gas density. 

Panel d in Figure \ref{fig:line} is the
Pa$\alpha$/H$_2$ ratio image which also traces the wind-blown cavity.
Note that the cone traced by Pa$\alpha$ and H$_2$ is offset with respect
to the light cone observed in R:
this is a result of the reddening gradient in the
direction perpendicular to the galactic plane.

Fig. \ref{fig:line}f is a true color RGB representation 
of line and continuum images (Red=F222M, Green=\Hmol, Blue=\PA).

L$^\prime$ and N band ground based images are shown in Fig. \ref{fig:midir}
a and b with contours overlayed on the NICMOS K band image.
No obvious point source is detected at the location of the nucleus
and the extended emission is smooth and regular, elongated as
the galactic disk. 

The \FeII\ emission shown in contours in Fig. \ref{fig:line}b deviates from the
\PA\ image in a number of interesting ways. First, the northern
edge of the cavity outlined most clearly in \Hmol\ emission is also detected, although more
faintly, in \FeII, presumably excited by the shocks resulting
from the superwind. Otherwise, the \FeII\ emission displays two
prominent peaks in the starburst region traced by \PA, one peak
close to the nucleus and one offset at a position angle of about 250
degrees (counterclockwise from North). In both of these regions the
\FeII/\PA\ ratio is much higher than in the rest of the starburst region. The
\FeII\ emission likely originates in radiative supernova remnants
(SNRs). In the dense nuclear region of NGC4945 the radiative phase of
the SNRs will be short, and hence the \FeII\ emission will be much more
strongly affected by the stochastic nature of supernova explosions in
the starburst ring than \PA. The regions of high
\FeII/\PA\ ratios thus simply trace recent supernova activity.

\subsection{Reddening}

A lower limit and a reasonable estimate of reddening can be
obtained from the H--K color image in the case of foreground screen extinction.
In this case, the extinction is simply
\begin{equation}
\AV = \frac{E(H-K)}{c(H)-c(K)}
\end{equation}
where the color excess is given by the difference between observed 
and intrinsic colour, $E(H-K)=(H-K)-(H-K)_\circ$ and
the $c$ coefficients represent the wavelength dependence of the extinction law;
$A_\lambda = c(\lambda) \AV$.
We have assumed $\AWL=A_{1\MIC}(\WL/1\MIC)^{-1.75}$
($\lambda>1\MIC$) and $\AV=2.42 A_{1\MIC}$.
Spiral and elliptical galaxies have average intrinsic colours
$(H-K)_\circ\sim 0.22$ with 0.1\MAG\ dispersion (Hunt et al. \cite{hunt})
and the color correction due to the non-standard filters used by NICMOS
is negligible -- $(H-K)_\circ\sim 0.26$ instead of 0.22.
In the region of the \PA\ ring,
the average color $H-K=1.1$ yields $\AV\simeq 11$,
in fair agreement with the estimate $\AV>13$ from the Balmer
decrement presented below. In knot B $(H-K)=1.2$ yields $\AV=12.5$, while
in knot C $(H-K)=0.64$ yields $\AV=5.2$. 

A different reddening estimate can be derived from the analysis
of Hydrogen line ratios. We can estimate the reddening to "Knot B" and "Knot C"
by using the images and spectra published by 
Moorwood et al. (\cite{moorwood96a}).
The inferred reddening (assuming an intrinsic ratio Pa$\alpha$/H$\alpha$=0.18,
and $A(\HA)=0.81\AV$, $A(\PA)=0.137\AV$) is \AV=3.2 for Knot B and
\AV=3.8 for Knot C.
We can also estimate a lower limit to the reddening on the 
\PA\ ring. Considering a region $\sim 11\arcsec\times5\arcsec$ aligned
along the galactic plane, including all the stronger \PA\ emission, 
we find $\PA/\HA>500$ which corresponds to $\AV>13$mag, a value
in agreement with the estimate given by Moorwood \& Oliva (\cite{moorwood88}),
$\AV=14\pm3$, from the \BA/\BG\ ratio in a $6\arcsec\times6\arcsec$
aperture centered on the IR peak.

We note that the first approach measures the mean extinction of the starlight,
while the second one measures the extinction toward the HII regions.
Therefore, these \AV\ estimates indicate that in the case of Knot C
the star light and the emitting gas are located behind the same screen.
Conversely, Knot B has a lower extinction and must therefore be
located in front of the screen hiding the star light.
A likely interpretation is that Knot C is located within the galactic 
plane on the walls of the cavity farthest from us.
whereas Knot B is 
located above the galactic plane, toward the observer
 
It appears that the hypothesis of screen extinction
can provide reasonable results.
Of course the true extinction, i.e. the optical depth at a
given wavelength, is larger if dust is mixed with the emitting regions.
However, it should be noted that the case in which dust is completely
and uniformly mixed with the emitting regions does not apply here
because the observed color excesses are larger than the maximum
value expected in that case ($E(H-K)\sim 0.6$).

\subsection{CO Index}

A straight computation of the CO stellar index as\\
$W(CO)=m(CO)-m(K)$,
where $m(CO)$ and $m(K)$ are the magnitudes in the
CO and K filters, is hindered by the high extinction
gradients.
Therefore we have corrected for the reddening using the 
prescription described above:
\begin{equation}
W(CO) = m(CO)-m(K)+\frac{c(K)-c(CO)}{c(H)-c(K)}E(H-K)
\end{equation}
where, as above, the $c$ coefficients represent the wavelength dependence of
the reddening law. The correction is $0.145\, E(H-K)$ which is important
since the expected CO index is $\sim 0.2$.

The "corrected" photometric CO index map is displayed in Fig. \ref{fig:line}e.

As a check, in the central $4\arcsec\times 4\arcsec$ we derive
a photometric CO index of 0.18 which is in good
agreement with the value 0.22 obtained from spectroscopic observations by
Oliva et al. (\cite{oliva95}), when one takes into account the uncertainties
of reddening correction.

In the central region there are three knots
where the CO index reaches values $\simeq 0.25$ aligned along
the galactic disk.
However, we do not detect 
any clear indication of dilution by a spatially unresolved source,
that would be expected in the case of emission by hot 
($\sim 1000\K$) dust heated by the AGN.
There are regions close to the location of the H$_2$O maser where
the CO index is as low as 0.08 but that value is still consistent
with pure stellar emission or, more likely, with an imperfect
reddening correction.

\subsection{\label{sec:AGNactiv}AGN activity}

The NICMOS observations presented in this paper
were aimed at detecting near-IR traces of AGN activity
in the central ($R<10\arcsec$) region of NGC 4945. Indeed, recent NICMOS studies exploiting the high spatial resolution of HST
show that active galactic nuclei are usually
characterized by prominent point sources in K, 
detected e.g. in the Seyfert 2 galaxies
Circinus (Maiolino et al. \cite{maiolino99}) 
and NGC 1068, and the radio galaxy
Centaurus A (Schreier et al. \cite{schreier98}). 
NGC 4945 does not show any point-like emission at the position of the nucleus
(identified by the \water\ maser) and the 
upper limit to the nuclear emission is
$F_\lambda(\mathrm{F222M})<2\xten{-13}\ERG\CM\2\S\1\MIC\1$.

We also do not detect any dilution of the
CO absorption features by hot dust emission, as observed in many 
active galaxies (Oliva et al. \cite{oliva99b}). 
From the analysis of the CO index image, non-stellar light contributes less
than $F_\lambda(\mathrm{F222M})<6\xten{-14}\ERG\CM\2\S\1\MIC\1$
thus providing a tighter upper limit than above.

The lack of a point source in the ground based L and N observations 
also places upper limits on the mid-IR emission, though less tight 
due to the lower sensitivity and spatial resolution
($F_\lambda(\mathrm{L})<1.2\xten{-12}\ERG\CM\2\S\1\MIC\1$ and
$F_\lambda(\mathrm{N})<6.0\xten{-13}\ERG\CM\2\S\1\MIC\1$).

Finally, type 2 AGNs are usually characterized by ionization cones 
detected either in line images or in 
excitation maps, i.e. ratios between high and
low excitation lines (usually \OIII\ and \HA) revealing
higher excitation than the surrounding medium.
In NGC 4945 the equivalent width of \PA\ and the \PA/\Hmol\ 
ratio indeed show a cone morphology but the behaviour is the
opposite of what expected, i.e. the excitation 
within the cone is lower than in the surroundings and
the H$_2$/Pa$\alpha$ ratio increases up to $\sim 5$ (see Fig. \ref{fig:line}d).
Two processes could be responsible for the enhanced \Hmol\ emission --
either shocks caused by the interaction between the supernova-driven wind
and the interstellar medium or exposure to a strong X-ray dominated
photon flux emitted by the AGN. But in any case there is absolutely no
indication of the strong UV flux which produces ``standard'' AGN ionization
cones.

We find, therefore, no evidence for the expected AGN markers in our NICMOS
data.

\section{\label{sec:discuss}Discussion}

Although no trace of its presence has been found in these data,
the existence of an obscured AGN in the nucleus of
NGC 4945 is unquestionably indicated by the
X-rays (Iwasawa et al. \cite{iwasawa93}, Done et al. \cite{done96}).
Recent, high signal-to-noise observations by BeppoSAX
(Guainazzi et al., \cite{guainazzi}) have confirmed the
previous indications of variability 
from Ginga observations (Iwasawa et al. \cite{iwasawa93}):
in the 13-200 keV band, where
the transmitted spectrum is observed, the light curve 
shows fluctuations with an
extrapolated doubling/halving time scale of $\tau\sim 3-5\xten{4}\S$.
These time scales and amplitudes essentially exclude any known
process for producing the high energy X-rays other than 
accretion onto a supermassive black hole.

Making the 3\xten{42}\ERG\S\1\ observed in the
2-10\KEV\ band with BeppoSAX would require about 10000
of the most luminous X-ray binaries observed in our Galaxy 
(e.g. Scorpio-X1) and only a few of this objects are known.

Alternatively, very hot plasma ($KT\sim$ a few \KEV), due to 
supernovae, has been observed in the 2-10\KEV\ spectrum of starburst
galaxies, but at higher
energies ($>30\KEV$) the emission is essentially negligible
(Cappi et al. \cite{cappi}; Persic et al. \cite{persic});
whereas the
emission of NGC4945 peaks between 30 and 100\KEV. Also, given that the X-ray
emission is observed through a gaseous absorbing column density
of a few times \ten{24}\CM\2, both
the 10000 superluminous X-ray binaries and the very hot SN wind should be
hidden by this huge gaseous column. It is very difficult to find a geometry for
the gas distribution that could produce this effect. We
therefore conclude that the presence of an AGN provides 
the only plausible origin of the hard X-ray emission.

The above considerations combined with the absence of any evidence for the
presence of an AGN at other wavelengths
has important consequences irrespective of the relative,
and unknown, contributions of the starburst and AGN to the total bolometric
luminosity.
This is illustrated below by considering the extreme possibilities that the
luminosity is dominated either by the starburst or the AGN.

\subsection{NGC 4945 as a starburst dominated object}

Most previous studies have concluded that the FIR emission in NGC 4945
can be attributed solely to starburst activity (e.g. Koornneef \cite{koorn},
Moorwood \& Oliva \cite{moorwood94a}) without invoking 
the presence of an AGN.

We note that, on average, active galaxies are characterized by 
$L_{FIR}/L_{Br\gamma}$ ratios much larger than starbursts
and this fact was sometimes invoked to discern starbursts
from AGNs (see the discussion in Genzel et al. \cite{genzel98}).
In this regard, NGC 4945 has a starburst-like ratio:
$\LFIR/L_\mathrm{Br\gamma}\sim1.4\xten{5}$
(from observed \PA\ with \AV=15mag). This is similar to the 
value for the prototypical starburst galaxy M82
($\LFIR/L_\mathrm{Br\gamma}\sim3.4\xten{5}$, Rieke et al. \cite{rieke80}),
suggesting that the FIR emission of NGC 4945
may arise from the starburst.

Genzel et al. (\cite{genzel98})
showed that, when considering the reddening
correction derived from the mid-IR -- usually much larger than from the optical
and near-IR -- the observed H line emission from the
starburst translates into an
ionizing luminosity comparable to the FIR luminosity.
Indeed, if in NGC 4945 the bulk of H emission is hidden by just \AV=45mag,
$\LFIR/L_\mathrm{ion}\sim 1$ and the observed starburst activity
is entirely responsible for the FIR.

\begin{figure*}
\centerline{
\epsfig{figure=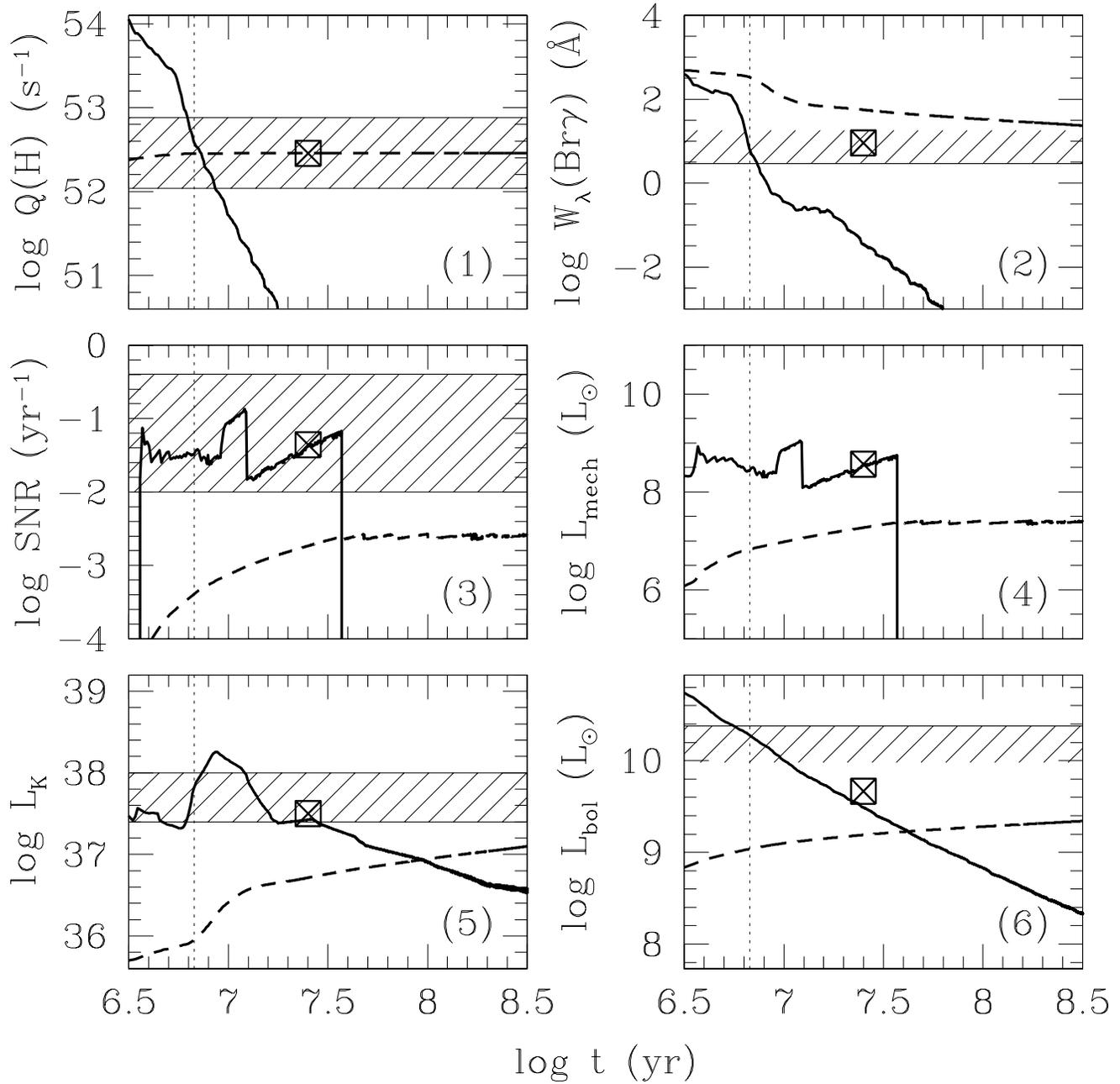,width=\linewidth}}
\caption{\label{fig:sbmod} Properties of a burst of star formation as a
function of the time elapsed from the beginning of the burst
(models by Leitherer et al. 1999).
The thick solid line represents an instantaneous burst with mass
$3.5\xten{7}\Mo$. The thick dashed line is a continuous star formation
rate of $0.13\Mo\YR\1$. See text for more details on the models.
Panel 1 is the time dependence of the ionizing photon rate. The shaded
area limits values consistent with observations. The thin dotted line is drawn 
at a time in which the "instantaneous" burst meets the observational
constraints. The crossed square represents the combination of the properties
of the two models at $t=10^{7.4}\YR$.
Panel 2 gives the \BG\ equivalent width. The shading outline the lower limit 
given by observations. The other symbols are as in panel 1.
Panel 3 gives the Supernova Rate. As above the shaded area limits
the range of values allowed from observations. Symbols as in panel 1.
Panel 4 gives the mechanical luminosity. Symbols as in panel 1.
Panel 5 gives the monochromatic luminosity
in the K band (\ERG\S\1\AA). Symbols as in panel 1.
Panel 6 gives the bolometric luminosity of the burst.
The shading marks the upper limit set by the total IRAS
luminosity of the galaxy.}
\end{figure*}

Although all the bolometric luminosity could be generated by a starburst 
it is also possible to construct starburst models which are consistent with the
observed near infrared properties but generate a much lower total luminosity.
It is important to recall that $\LFIR/L_{\BG}$
represents the ratio between star formation rates
averaged over two different timescales, i.e. $>10^8$yrs and $<10^7$yrs,
respectively. Therefore, this ratio strongly depends on the 
past star formation history. For example, objects which have not
experienced star formation in the past $10^7$yrs will emit
little \BG, but significant FIR radiation.
A more quantitative approach is presented in
Fig. \ref{fig:sbmod} where we compare the observed nuclear properties
of NGC 4945 with 
synthesis models by Leitherer et al. (\cite{leitherer}).
We have considered two extreme cases of star formation history.
The thick solid line in the figure represents an instantaneous burst 
with mass $3.5\xten{7}\Mo$ whereas the thick dashed line is a continuous 
star formation rate of $0.13\Mo\YR\1$.
In both cases a Salpeter initial mass function (i.e. $\propto M^{-2.35}$),
upper mass cutoff of 100\Mo\ and abundances $Z=\Zo$ are chosen.
Panel 1 shows the evolution of the ionizing photon rate
(\QH) as a function of time after the beginning of the burst.
The shaded region limits the values compatible with the observations;
\QH\ is estimated from the total \PA\ flux in the 
NICMOS images ($5.6\xten{-13}\ERG\S\1\CM\2$),
dereddened with $A_V=5$mag and $A_V=20$mag and converted 
using case B approximation for H recombinations.
Panel 2 gives the equivalent width
of \BG\ ($W_\lambda(\BG)$); the observed value
outlined by the shaded area is a lower limit
for the starburst models and was derived by rescaling the observed \PA\
flux and dividing by the flux observed in the same aperture with the F222M
filter.
Panel 3 is the evolution of the SuperNova Rate (SNR).
Estimates of SNR from radio
observations suggest values $>0.3\YR\1$ (Koornneef \cite{koorn}), $0.2\YR\1$   
(Forbes \& Norris \cite{forbes}),
down to 0.05\YR\1 (Moorwood \& Oliva \cite{moorwood94a}). 
The shaded region covers the 0.01-0.4\YR\1\ range.
Panel 4 is the mechanical luminosity produced by the Supernovae.
Finally, panels 5 and 6 give the K-band and
bolometric luminosity, respectively.
The allowed range for the K monochromatic luminosity
is given by the total observed flux in a $6\arcsec\times 6\arcsec$ aperture
centered on the K peak where photospheric emission from supergiants
is known to dominate (Oliva et al. \cite{oliva99b}).
The upper and lower limits represent
the values obtained after dereddening by $A_V=5$mag and $A_V=20$mag.
The upper limit to the bolometric
luminosity is the {\it total} NGC 4945 luminosity derived from IRAS observations
(Rice et al. \cite{rice88}).
In all cases the thin dotted line represents the time at which the properties
of the instantaneous burst meet the observational constraints.
The crossed square represent the combination of the two models
at $t=\ten{7.4}\YR$.

It is clear from the figure that an instantaneous burst of $t\sim\ten{6.8}\YR$
is capable of meeting all the observational constraints. It reproduces the
correct supernova rate and K band luminosity and its bolometric luminosity
dominates the total bolometric luminosity of the galaxy.
Conversely the continuous burst fails to reproduce the SNR and K luminosity.
Just considering these two models alone it is tempting to infer
that the starburst
powers the bolometric emission of NGC4945.
However, the instantaneous and continuous SFR are two extreme and simplistic
cases. More realistically the SF history is more complex since bursts
have a finite and limited length or are the combination of several 
different events. As an example we consider
the case of two bursts of star formation taking place at
the same time: one instantaneous and the other continuous. Both have the
same characteristics as the bursts presented above.
The properties of this double burst model at $t=\ten{7.4}\YR$ are shown 
in the figure by the crossed squares. The choice of the time
is arbitrary and any other value between $\ten{7.2}\YR$ and $\ten{7.5}\YR$
might do.
Even in this case the starburst model meets all the observational
constraints: \QH\ is provided for by the continuous burst while
SNR and K luminosity come from the instantaneous burst.
The important difference with respect to the single instantaneous
burst is that the bolometric luminosity of the burst is
now $\lesssim 20\%$ of the total bolometric luminosity of the galaxy.

The mechanical luminosity injected by the SN in the "instantaneous"
burst (which dominates also in the double burst model)
is $\sim \ten{8.5}\Lo$ over $\sim \ten{7.4}\YR$. This results
in a total injected energy of $\sim 10^{57}\ERG$ which is more than enough
to account for the observed superwind.
Indeed Heckman et al. (\cite{heckman}) estimate
an energy content of the winds
blown cavity of $\sim 1.5\xten{55}\ERG$
(after rescaling for the different adopted distance of NGC 4945).
Both models agree with the constraints imposed by dynamical measurements 
that the central mass in stars must be less than 6.6\xten{8}\Mo\
(Koornneef \cite{koorn}, after rescaling for the different assumed distances of
the galaxy): the continuous SFR would require 5\xten{9}\YR\
to produce that mass of stars.

In conclusion, two different star formation histories can reproduce
the observed starburst properties but only in one case
does the starburst dominate the bolometric luminosity of the galaxy.
Therefore the available data
do not allow any constraints on the bolometric luminosity of the starburst.

As shown in the next section, the observed (\LFIR/\LX) ratio
of NGC 4945 is equal to that of a "normal" AGN in which the
\LFIR\ is reprocessed UV radiation. If the \LFIR\ in NGC 4945 is actually 
dominated by the starburst, therefore, it is clear that the AGN must
be strongly deficient in UV relative to X-rays.

In this starburst-dominant scenario for NGC4945, 
with the black hole mass inferred from the \water\ maser
measurements (Greenhill. et al. \cite{greenhill}), the AGN is emitting
at \lesssim 10\% of its Eddington Luminosity.

\subsection{NGC 4945 as an AGN dominated object}

\begin{figure}
\epsfig{figure=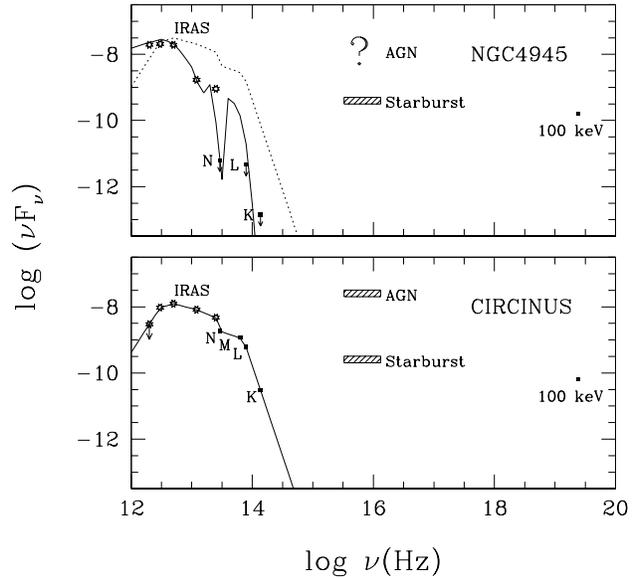,width=\linewidth}
\caption{\label{fig:compare}
Spectral energy distributions of NGC 4945 (upper panel) and
Circinus (lower panel). "Stars" are the IRAS
photometric points (except for the points with the highest wavelength
which are from baloon-borne observations). "K", "L", "M" and "N"
are the points in the standard photometric bands. The hatched areas
labeled as "Starburst" and "AGN" represent the continuum levels
derived from the ionizing photon rates (see text) emitted by starburst 
and AGN, respectively. The "100\KEV" points are from X-ray observations.
The IR spectrum of Circinus (solid line in lower panel) is plotted in the
upper panel (dotted line) after rescaling to match the 100\KEV\ points.
The solid line in the upper panel is the same spectrum after extinction
by an extra $\AV=150\MAG$ (see text for details). }
\end{figure}

By fitting the simultaneous 0.1-200 keV spectrum from BeppoSAX,
the absorption corrected luminosity in the 2-10 keV band
is $\LX(2-10\KEV) = 3\xten{42}\ERG\S\1$ (Guainazzi et al., \cite{guainazzi}).
If the AGN in NGC 4945 has an intrinsic spectral energy distribution
similar to a quasar, then
$\LX(2-10\KEV)/L_{\rm bol}\sim 0.03$ (Elvis et al. \cite{elvis}) therefore
$(L_{\rm bol})_{\rm AGN}\sim \ten{44}\ERG\S\1 = 2.6\xten{10}\Lo$
which is the {\it total} far-IR luminosity of NGC 4945, measured by IRAS
(Rice et al., \cite{rice88}).
Thus, a "normal" AGN in NGC 4945 {\it could} in principle
power the total bolometric luminosity.

For this scenario, we compare NGC 4945 with a nearby obscured object,
the Circinus galaxy,
now considered an example of a ``standard'' Seyfert 2 galaxy
(c.f. Oliva et al. \cite{oliva94}, Oliva et al. \cite{oliva98}, 
Maiolino et al. \cite{maiolino98}, Matt et al. \cite{matt99},
Storchi-Bergmann et al. \cite{storchi99}, Curran et al. \cite{curran}).
In particular, Oliva, Marconi \& Moorwood (\cite{oliva99}) and,
previously, Moorwood et al. (\cite{moorwood96c})
showed that the total energy output from the AGN required to 
explain the observed emission line spectrum is comparable to the
total FIR luminosity, concluding that any starburst contribution
to the bolometric luminosity is small (\lesssim 10\%).
The choice of the Circinus galaxy is motivated by the 
similar distance (D=4\MPC), FIR and hard X-ray luminosities as NGC 4945
($\LFIR\sim1.2\xten{10}\Lo$; Siebenmorgen et al., \cite{sieben} --
$\LX(2-10\KEV)\sim 3.4-17\xten{41}\ERG\S\1$; Matt et al., \cite{matt99}).
Note that its $\LX (2-10\KEV)/\LFIR$ ratio ($\sim 0.01-0.05$)
is consistent with the average value for quasars (Elvis et al., \cite{elvis}).

The overall spectral energy distributions of NGC 4945 and Circinus
are compared in Fig. \ref{fig:compare}.
The "stars" represent the IRAS photometric points (except for the
points with the largest wavelength which are the 150\MIC\
measurements by Ghosh et al. \cite{ghosh92}).
In NGC 4945 the points labeled with "K", "L" and "N" are the upper limits
derived from our observations, while in Circinus they represent emission from
the unresolved nuclear source corrected for stellar emission
(Maiolino et al. \cite{maiolino98}).
The points labeled "100 keV" are from Done et al. \cite{done96} (NGC 4945)
and Matt et al. \cite{matt99} (Circinus).
The bars between 13.6 and 54.4 eV are at a level given by
$\nu L_\nu\sim Q(\mathrm{H}) <h\nu>$, where $Q(\mathrm{H})$ is the rate
of H-ionizing photons and $<h\nu>$ is the mean photon energy of the ionizing
spectrum. For NGC 4945, $Q(\mathrm{H})$ is derived from H recombination 
lines and thus represent 
the energy which is radiated by the young starburst; 
we assumed $<h\nu>=16\EV$.
For Circinus, the point labeled with "Starburst" is similarly derived
from Br$\gamma$ emission associated with the starburst
(Oliva et al. \cite{oliva94}) while that labeled "AGN"
is from the estimate made by Oliva, Marconi \& Moorwood (\cite{oliva99}).

In the lower panel, we represent the IR spectrum of Circinus by connecting
the photometric points just described. We plot this same spectrum as a dotted line
in the upper panel, rescaling to match the
100 keV points. NGC 4945 and Circinus have similar X/FIR ratios:\\
$\nu L_\nu(100 keV) / L_{FIR} \simeq 2\xten{-3}$ for Circinus
and $\simeq 3\xten{-3}$ for NGC 4945.
Note that, at each wavelength, both Circinus and NGC 4945 were observed
with comparable resolution.

If the AGN in NGC 4945 dominates the luminosity and its intrinsic spectrum
is similar to that of Circinus, then the lack of AGN detections in the near-IR
and mid-IR require larger obscuration.
In particular the non-detection of a K band point source or of dilution
of the CO features can be used to estimate the extinction at 2.2\MIC.
In Circinus from Maiolino et al. \cite{maiolino98} (K band) and
Matt et al. \cite{matt99},
we can derive $\nu L_\nu (K) / \nu L_\nu (100 keV)\simeq 0.5$.
If NGC 4945 has a similar near-IR over hard-X-rays ratio then,
given $\nu L_\nu (K) / \nu L_\nu (100 keV) < 8\xten{-4}$
(K band upper limit from CO and X-ray data from
Done et al. \cite{done96}), 
the extinction toward the nucleus is
$\Delta \mathrm{A}_\mathrm{K}>7$mag (i.e. $\Delta\AV>70$mag) 
larger than in the case of Circinus.
Hot dust in NGC 4945 must be hidden by at least 
$\AV>135\MAG$, in agreement with the estimate by
Moorwood \& Glass (\cite{moorwood84}),
$\AV>70$ and, more recently, with an analysis of ISO CVF spectra
implying $\AV\sim 100\MAG$ (Maiolino et al., 2000, in preparation).
We note that the required extinction is not unexpected
and in agreement with the X-ray measurements.
The measured column density in absorption in the X-rays is 
$N_\mathrm{H}\sim {\rm few}\xten{24}\CM\2$ therefore the expected $A_V$,
assuming a galactic gas-to-dust ratio is:
\begin{equation}
A_V\sim 450 \left( \frac{N_\mathrm{H}}{\ten{24}\CM\2} \right)
\end{equation}
The $A_V$ measured from optical/IR data is estimated smaller
than derived from X-rays ($A_V(\mathrm{IR})\sim 0.1-0.5\, A_V(\mathrm{X})$;
Granato, Danese \& Franceschini \cite{granato}),
therefore the X-ray absorbing
column density is in excellent agreement with the required extinction.
Very high extinction, expected in the frame of the unified AGN model
are observed in many objects as discussed and summarized,
for instance, in Maiolino et al. \cite{maiolino98b}
and in Risaliti et al. \cite{risaliti}.  

The higher extinction can also qualitatively explain the redder
colors of the NGC 4945 FIR spectrum.
The solid line
in the upper panel is the spectrum of Circinus after applying
foreground extinction
by $\AV=150\MAG$. We have applied the extinction law by Draine
\& Lee (\cite{draine}) and the energy lost in the mid-IR has been reprocessed
as 40\K\ dust emission  (i.e. black body emission at 40\K\ corrected
for $\WL^{-1.75}$ emissivity).
Though a careful treatment requires a full radiation transfer calculation,
this simple plot demonstrates that (i) the redder color of NGC 4945 with
respect to Circinus can be explained with extra absorption and (ii) that 
this is not energetically incompatible with the observed FIR luminosity,
i.e. the absorbed mid-IR emission re-radiated
in the FIR does not exceed the observed points.

If the FIR emission is powered by the AGN this is UV radiation re-processed
by dust. However, if the AGN emits $\sim 2\xten{10}\Lo$ in UV photons,
high excitation gas emission lines should also be observed.
The absence of high ionization lines like
\OIII$\lambda 5007,4959$\AA\ (Moorwood et al. \cite{moorwood96a})
or \NeV$\lambda 14.3\MIC$ (Genzel et al. \cite{genzel98}) 
and the low excitation observed in the wind-blown cone strongly
argues that no ionizing UV photons
(i.e. $13.6\le \HNU < 500\EV$) escape from the inner
region.
The low excitation H$_2$/Pa$\alpha$ map,
associated with the peak in H$_2$
emission close to the nucleus location, indicates that 
ALL ultraviolet photons must be absorbed within
$R<1\farcs5$, i.e. $R<30\PC$ along ALL lines of sight.
This is in contrast with the standard unified model of AGN
where ionizing radiation escapes
along directions close to the torus axis.

If the AGN is embedded in a thick dusty medium then two
effects will contribute to its obscuration. First,   
dust will compete with the gas in absorbing UV photons
which will be directly converted into infrared radiation
(e.g. Netzer \& Laor \cite{netzer93},
Oliva, Marconi \& Moorwood \cite{oliva99}).  Second,
emission lines originating in this medium
will be suppressed by dust absorption. To estimate the amount of required
extinction, note that in Circinus $\NeV 14.3\MIC/\NeII 12.8\MIC=0.4$
(extinction corrected) and in NGC 4945 \NeV/\NeII$\le 0.008$
(both ratios are from Genzel et al. \cite{genzel98}).
If NGC4945 has the same intrinsic ratio as Circinus, then the observed
\NeV/\NeII\ ratio requires $A(14.3\MIC)>4.2$mag  
corresponding to $\AV>110$mag and in agreement
with the above estimates.

We conclude that the AGN can power the FIR emission if it is properly obscured.  Inferring
the black hole mass from the \water\ maser
observations ($1.4\xten{6}\Mo$, Greenhill. et al. \cite{greenhill}), we find in this scenario that
the AGN is emitting
at $\sim 50\%$ of its Eddington Luminosity.

\subsection{On the existence of completely hidden Active Galactic Nuclei}

As discussed above, if an AGN powers the FIR
emission of NGC 4945, it must be hidden up to mid-IR 
wavelengths and does not fit in the standard unified model.  
The possible existence of such a class of Active Nuclei,
detectable only at $>10\KEV$,
would have important consequences on the interpretation of
IR luminous objects whose power source is still debated.

Genzel et al. (\cite{genzel98}) and Lutz et al. (\cite{lutz98}) compared
mid-IR spectra of Ultra Luminous IRAS galaxies (ULIRGs, see
Sanders \& Mirabel \cite{sanders96} for a review) with those of AGN and
starburst templates. They concluded that the absence of high excitation lines
(e.g. \NeV) and the presence of PAH features
undiluted by strong thermal continuum in ULIRGs spectra
strongly suggest that the starburst component is dominant.
They also show that, after a proper extinction correction,
the observed star formation activity can power FIR emission.
In their papers, NGC 4945 is classified as a starburst 
because of its mid-IR properties but, as shown in the previous section, 
NGC 4945 could also be powered by a highly obscured AGN
and the same scenario could in principle apply to all ULIRGs.
Their bolometric emission can be powered by an active nucleus
completely obscured even at mid-IR wavelengths. 

The same argument could be used for 
the sources detected at submillimeter wavelengths by SCUBA
which can be considered
as the high redshift counterpart 
of local ULIRGs. If they are powered by hidden active nuclei then
their enormous FIR emission would not require star formation
rates in excess of $>100\Mo\YR\1$ (e.g. Hughes et al. \cite{hughes98}),
and this would have important consequences for understanding the history
of star formation in high redshift galaxies.

In addition, it is well known that in order to explain the X-ray background
a large fraction of obscured AGN is required. However
Gilli et al. (\cite{gilli99}) have shown that, in order to reconcile
the observed X-ray background with hard X-ray counts, a rapidly
evolving population of hard X-ray sources is required
up to redshift $\sim 1.5$. No such population is
known at the moment and the only class of objects which are known
to undergo such a rapid density evolution are local ULIRGs
(Kim et al. \cite{kim98}) and, at higher redshift,
the SCUBA sources (Smail et al. \cite{smail97}).
SCUBA sources are therefore candidates to host a population
of highly obscured AGNs.

Almaini et al. (\cite{almaini99}) suggest that, if the SED of high redshift
AGN is similar to those observed locally, one can explain
10--20\% of the 850\MIC\ SCUBA sources at 1 mJy.
This fraction could be significantly higher if a large
population of AGN are Compton thick at X-ray wavelengths.
Trentham, Blain \& Goldader (\cite{trentham}) show that if the SCUBA sources
are completely powered by a dust enshrouded AGN then they may
help in explaining the discrepancy between the local density in 
super massive black holes and the high redshift AGN component
(see also Fabian \& Iwasawa \cite{fabian99}).  

Establishing the nature of SCUBA sources could be extremely
difficult if the embedded AGNs are like NGC4945, i.e.
completely obscured in all directions, because they would then not be
identifiable with the standard optical/IR diagnostics.
Incidentally, this fact could possibly account for the sparse detections
of type 2 AGNs at high redshifts (Akiyama et al. \cite{akiyama}).

The best possibility for the detection of NGC4945-like AGNs is 
via their hard X-ray emission but, unfortunately,
the sensitivity of existing X-ray surveys is still not high enough 
to detect high $z$ AGN and the low spatial resolution makes identifications
uncertain in the case of faint optical/near-IR counterparts.
Moreover, hard X-rays alone are not enough to establish if the AGN
dominates the bolometric emission.

\section{\label{sec:conclus}Conclusions}

Our new HST NICMOS observations of NGC 4945, complemented by new ground
based near and mid-IR observations, have provided detailed morphology
of the nuclear region. In \PA, we detect a 100pc-scale starburst ring while
in \Hmol\ we trace the walls of a conical cavity blown by
supernova driven winds. The continuum images are strongly affected by
dust extinction but show that even at HST resolution and sensitivity,
the nucleus is completely obscured by a dust lane with an elongated, disk-like
morphology.
We detect neither a strong point source nor dilution in CO 
stellar features, expected signs of AGN activity.

Whereas all the infrared properties of NGC 4945 are consistent with starburst
activity, its strong and variable hard X-ray emission cannot be plausibly 
accounted for without the presence also of an AGN.
Although the starburst must contribute to the total bolometric luminosity
we have shown, using starburst models, that the actual amount is dependent on the star formation history. A major contribution from the AGN is thus not
excluded. Irrespective of the assumption made, however, our most important 
conclusion is that the observed variable hard X-ray emission combined
with the lack of evidence for reprocessed UV radiation in the
infrared is incompatible with the "standard" AGN model. If the AGN dominates
the bolometric luminosity, then its UV radiation must be totally obscured along all lines of sight. If the starburst dominates then the AGN must be 
highly deficient in its UV relative to X-ray emission. The former case clearly raises the possibility that a larger fraction of ULIRGs than currently
thought could actually be AGN rather than starburst powered.

\begin{acknowledgements}
A.M. and A.R. acknowledge the partial support of the Italian Space Agency (ASI)
through grants ARS--98--116/22 and ARS--99--44
and of the Italian Ministry for University
and Research (MURST) under grant Cofin98-02-32.
E.J.S. acknowledge support from STScI GO grant O0113.GO-7865.
We thank Roeland P. van der Marel for the use of the pedestal
estimation and quadrant equalization software.
\end{acknowledgements}

\end{document}